\documentstyle[aps,twocolumn]{revtex}

\begin{document}
\draft
\author{Bao-Sen Shi and Akihisa Tomita}
\address{Imai Quantum Computation and Information Project, NEC Tsukuba Laboratories,\\
ERATO, Japan Science and Technology Corportation (JST)\\
Fundamental Research Laboratories, NEC, 34 Miyukigaoka, Tsukuba, Ibaraki,\\
305-8501, Japan}
\title{Generation of pulsed polarization entangled photon pair via a space cascaded
two-crystal geometry}
\maketitle

\begin{abstract}
The generation of pulsed polarization entangled photon pair has been
realized using type-I phase matching in the spontaneous parametirc
downconversion process in a space cascaded two-crystal geometry. The optical
axes of the crystal are aligned in such a way that the horizontal
polarization photon pair is produced from up crystal, and the vertical
polarization photon pair is produced from down crystal. These two processes
are simultaneous, but are separable in space. We get the high entangled
polarization photon pair by using the single mode fiber to erase the spatial
information between these two processes. This photon pair exhibits more than
86\% high-visibility quantum interference for polarization variable without
the narrowband filter and temporal compensation.
\end{abstract}


Quantum entanglement is one of the most striking features of quantum
mechanics. Entangled states of two or more particles not only provide
possibilities to test basic concepts of quantum mechanics[1], but also form
the basis of quantum information, make such phenomena as quantum
cryptography [2], teleportation [3], dense coding [4] and quantum
computation [5] possible. So, it is clear that preparation of maximally
entangled state , or Bell state , is an important subject in modern
experiment quantum optics. At present, the most accessible and controllable
source of entanglement is the process of Spontaneous Parametric
Down-Conversion (SPDC) in a nonlinear crystal. [6-10] SPDC can be used to
generate the entangled photon states using type-I or type-II phase matching.
In type-I phase matching, the two paired down-converted photons have the
same polarizations and they emitted along concentric cones around the
direction of pump beam. In type-II phase matching, the down-converted
photons have orthogonal polarizations. Those differently polarized photons
are emitted along two different cones with cone axes at opposite sides of
the pump beam. For two particular emission directions, the correlated
photons are produced in the state $\left| HV\right\rangle +\left|
VH\right\rangle $ [10]. Recently, Kwiat et. al used two thin nonlinear
crystals to prepare Bell state using non-collinear type-I SPDC. [11] In
their scheme, these two crystals were arranged in cascade with their optical
axes perpendicular to each other. By pumping with two orthogonally polarized
pump beams , one can generated , for example, horizontal photons from the
first crystal or vertical photons from the second crystals. With cw pump,
these two down-conversions are mutually coherent and form an entangled
state. Another kind of entangled state called ''Time-Bin'' entangled was
discussed by Gisin's group [12]. In cw pumped SPDC, there are readily
available well-developed methods of preparing the polarization Bell state.
For applications in quantum information field, cw pumped SPDC is not very
useful, because the entangled photon pairs occur randomly with the coherence
length of the pump laser. This huge time uncertainty makes it difficult to
use in many applications, such as quantum teleportation, quantum swapping,
the generation of multi-photon entangled state [13] and so on, because in
these processes, the interaction between different photons generated from
different sources are required. One way to solve this difficulty is using a
femtosecond pulsed laser as a pump. Unfortunately, femtosecond pulse pumped
type-II SPDC shows very poor quantum correlation compared to cw case due to
the very different behavior of two-photon wavepacket [14]. In order to get
high visibility of quantum interference, usually, two ways are used: one is
using very thin nonlinear crystal; another way is using a very narrow
interference filter in front of detector. But even we use one of these two
ways or both, we also can not achieve high interference visibility in
principle [14], besides reduce the available flux of entangled photon pair
significantly.

Recently, Kim et. al present a way for preparation of Bell state using
femtosecond pulse pumped SPDC. [15, 16] In their scheme, an interferometric
technique is used to remove the intrinsic distinguishability in
femtosecond-pulse-pumped SPDC process. The two photon entangled state
exhibits high-visibility quantum interference. The advantages of this method
is: visibility is insensitive to the thick of crystals, also insensitive to
the narrow-band filter. Usami et. al [17]used the same crystal geometry as
Kwiat's, but pumped by femtosecond pulse laser to produce entangled
polarization photon pair. In order to restore the high entanglement, they
used a narrow band interference filter in their experiment. Nambu et. al
[18] used the same crystal but used pre-compensation method to get high
entanglement. Bitton et. al [19] also present a scheme to produce the
polarization-entangled photon pair pumped by femtosecond laser. The geometry
of crystal is also similar to Kwiat's geometry. The difference is two
cascaded crystals are cut with type-II phase matching. Very recently, Kim
and Grice [20] give a theoretical method to generate the pulsed polarization
entangled photon pair via temporal and spectral engineering. In schemes [15,
16, 18, 19], a suitable optical delay should be included. If without this
delay, no Bell state can be produced. In Ref. [17], a very narrow band
interference filter is needed in order to get good quantum correlation. In
this paper, we present a novel way to produce pulsed polarization entangled
photon pair using type-I phase matching SPDC process in a space cascaded
two-crystal geometry, this polarization entangled photon pair exhibits
high-visibility quantum inference for polarization experimentally. The main
advantages of this scheme are as follows: we do not need to consider the
suitable time compensation; In principle, visibility should be insensitive
to the thickness of crystal, so we can use thicker crystal to increase the
intensity of photon pair; The post-selection on spectrum by the narrow
bandwidth filter is not needed. In the follows, we show our scheme in detail.

In our two-crystal geometry, two identical 5x2.5x1mm BBO crystals (produced
by CASIX optronics Inc.) are used. They are oriented with their optical axes
aligned in perpendicular plane, and stacked in vertical direction. Please
refer to Fig. 1. These two crystals are cut collinearly in type-I phase
matching, phase matching angle $\Theta _m=29.18^{\circ }$. If one beam with
diameter a and 45 degree polarization pumps these two crystals , then for
example, up crystal will produce horizontal polarization photon pair $\left|
HH\right\rangle $, and down crystal will produce vertical polarization
photon pair $\left| VV\right\rangle $. Obviously, these two processes are
simultaneous, but are separable in space. If we make the spatial information
about these two possible SPDC process indistinguishable, (In our experiment,
we use the single mode fiber to realize it.) then these two possible
downconversion processes are coherent. In our two-crystal geometry, we just
consider the collinear process, so a postselection on amplitude is needed in
order to get the polarization entangled state $\left| HH\right\rangle
+\left| VV\right\rangle $. (But postselection is not needed in principle,
for example, we can use non-degenerated phase matching to solve it)

Observation of high-visibility quantum interference is a test of the degree
of quantum entanglement. In quantum interference experiment for
polarization, more than 86\% visibility of two-photon quantum interference
is obtained without narrow band filter and any time compensation. Consider
the experimental setup shown in Fig. 2. A femtosecond laser from a
mode-locked Ti: sapphire laser (coherent: Vitesse) is used to pump a 1mm BBO
crystal cut with type-I phase matching to get frequency doubled radiation.
The wavelength of femtosecond laser is 800nm, the pulse width is less than
100fs. Power of laser is about 870mw. Repetition rate is 80MHz. A prism ,
two dichroic mirrors which transmit 800nm light and reflect 400nm light ,
one 40nm interference filter with center wavelength 400nm and a polarizer
are combined to cut remainder 800nm light. After pass a half waveplate,
about 70mw 400nm laser with 45 degree polarization is used to pump our
two-crystal SPDC source. We use another two dichroic mirrors and one prism
to cut remainder 400nm light. After that, we couple photon pair to a 1m long
single mode fiber (operation wavelength 800nm). The output of fiber is input
to a 50/50 beamsplitter. At each output port of the beamsplitter, a detector
package consisting of a Glan-Thompson analyzer A$_1$ or A$_2$ and a
single-photon detector (PerkinElmer SPCM-AQR-14-FC) are placed. We do not
use any narrow interference filter before detector, just place a bandpass
filter (CVI company, LPF-750-1.00, transmit from 750nm) before the
beamsplitter. The outputs of detectors are sent to coincidence circuit for
coincidence counting. The coincidence circuit consists of a
time-to-amplitude converter and single-channel analyzer (TAC/SCA) and a
counter. The time window of coincidence counting is 2ns. The key of this
experiment is how to make coincidence counting producing from two crystals
equal. In experiment, we do it by adjusting the height of crystal. Then we
fix the polarizer A$_2$ in 45 degree, and rotate the polarizer A$_1$ with
each step 10 degree. From theory, we know, for the state $\left|
HH\right\rangle +\left| VV\right\rangle $, it should exhibit the
polarization interference in coincidence counting rate if we do the
two-photon interference experiment, and interference pattern should
correspond to the expression: $R_c\propto \cos ^2(\Theta _1+\Theta _2)$,
where, $\Theta _{}$ and $\Theta _{}$ are angles of polarizer A$_1$ and A$_2$%
. Experimentally, we get the polarization interference pattern shown in Fig.
3. The visibility is more than 86\% (including accidental coincidence
counting). In principle, visibility should be 100\%. The reason that it does
not reach 100\% , we think, it mainly is that the photon pairs from
different crystals are not equal exactly. We can make them equal in
principle, but it is difficult during the experiment. From the Fig. 3, we
see that the first peak is lower than the second peak. We try experiment
several times, this phenomena always exists, so we think that it may be
caused by the small position shift of polarizer when it is rotated. During
experiment, the single counting is almost unchanged. We measure that the
fluctuation of single counting is less than 7\%. This small change maybe due
to expectedly high background light, for example, from BBO crystals. In our
experiment, background count is almost equal to true single count from SPDC
[21]. Such high background light mainly consists of two parts: one is from
bad dark condition of our laboratory; another is from light reflected from
many optical components, such as prisms, BBO crystals, mirrors etc. To make
sure the state is really a polarization Bell state, we change the polarizer A%
$_2$ to different angles and do the same experiments. Experimentally,
interference patterns also shift the same angle as polarizer A$_2$ shifts
and almost same visibilities of quantum interference are observed, which
means that the state we observed is truly a polarization entangled state.

One of main advantage of this scheme is that no any temporal compensation is
needed. In our setup, two pumps arrive at crystal at the same time. In each
crystal, the SPDC light is ordinary light, so no delay is introduced between
two SPDC processes if two crystals have same length. The photon pair from up
or down crystal leaves the crystal at the same time averagely, they are just
separable in space. So from temporal point, we do not need to do any
compensation like in previous experiments [15-19] using optical delay or
narrow band interference filter. Another advantage is that our scheme is
insensitive to length of crystal. The wavepackets of photon pairs from up
crystal and from down crystals are the same irrespective of crystal
lengthen. So, we can increase photon pair flux by using thicker crystal. The
disadvantage of this scheme is that it need erase the space distinguishable
information between two possible SPDC processes. In our experiment, we use
the degenerate collinear phase-matching, so it need amplitude postselection
to get the entangled state. But in principle, we can overcome this problem,
for example, using collinear non-degenerate phase-matching.

The other three Bell states can be prepared by inserting the combination of
a half waveplate and a $\pi $ phase shifter. Using our scheme,
non-maximally-entangled state , i.e., states of the form $HH+\epsilon VV,$%
where, $\left| \epsilon \right| \neq 1$, may be produced, simply by
adjusting the height of crystals or the angle of half waveplate. This kind
of state has been shown useful in reducing the required detector
efficiencies in loop-free tests of Bell's inequalities [22]. Moreover, by
our scheme, the arbitrary (partially) mixed state of type $\cos ^2\Theta
\left| HH\right\rangle \left\langle HH\right| +\sin ^2\Theta \left|
VV\right\rangle \left\langle VV\right| $ can be produced. We need only do no
compensation in space (for mixed state ) or partially compensate in space
(for partially mixed state).

In summary. We present a novel scheme to produce Bell state using
femtosecond pulse pump. We also show strong polarization correlation
experimentally. One of main advantage is that no optical delay or narrow
band interference filter for high quantum interference visibility. Besides,
we can use thick crystal to get high intensity of photon pair. The
disadvantage of this scheme is that the spatial compensation is needed.

We thank Prof. Imai for support. We thank K. Usami , Dr. S. Kouno for
valuable discussions and technical support. Shi also thank Dr. Y. K. Jiang
of Universit$\stackrel{`}{a}$ dell' Insubria for careful reading this
manuscript and useful comments.


\begin{references}
\bibitem{}  Because of high background light and colinear phase matching, it
is very difficult to get exact single counting from photon pairs. We just do
a raw estimation. We measure the single counting when we use a black plate
to block SPDC light and regard this value as a raw estimation of background
light. We use the value without black plate to subtract the value with the
plate, and regard this difference as the raw single counting estimation from
photon pairs. In our experiment, raw estimations about single counting from
photon pairs and from background are almost equal, both are about 1.6kc/s.
Based on these estimation, the accident coincidence counting is less than
4c/100s.
\end{references}
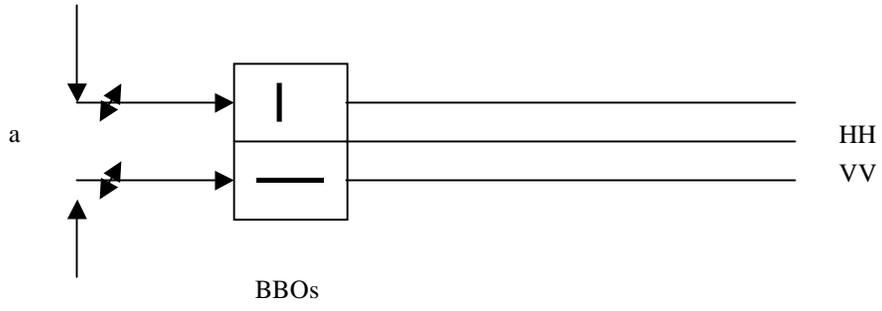
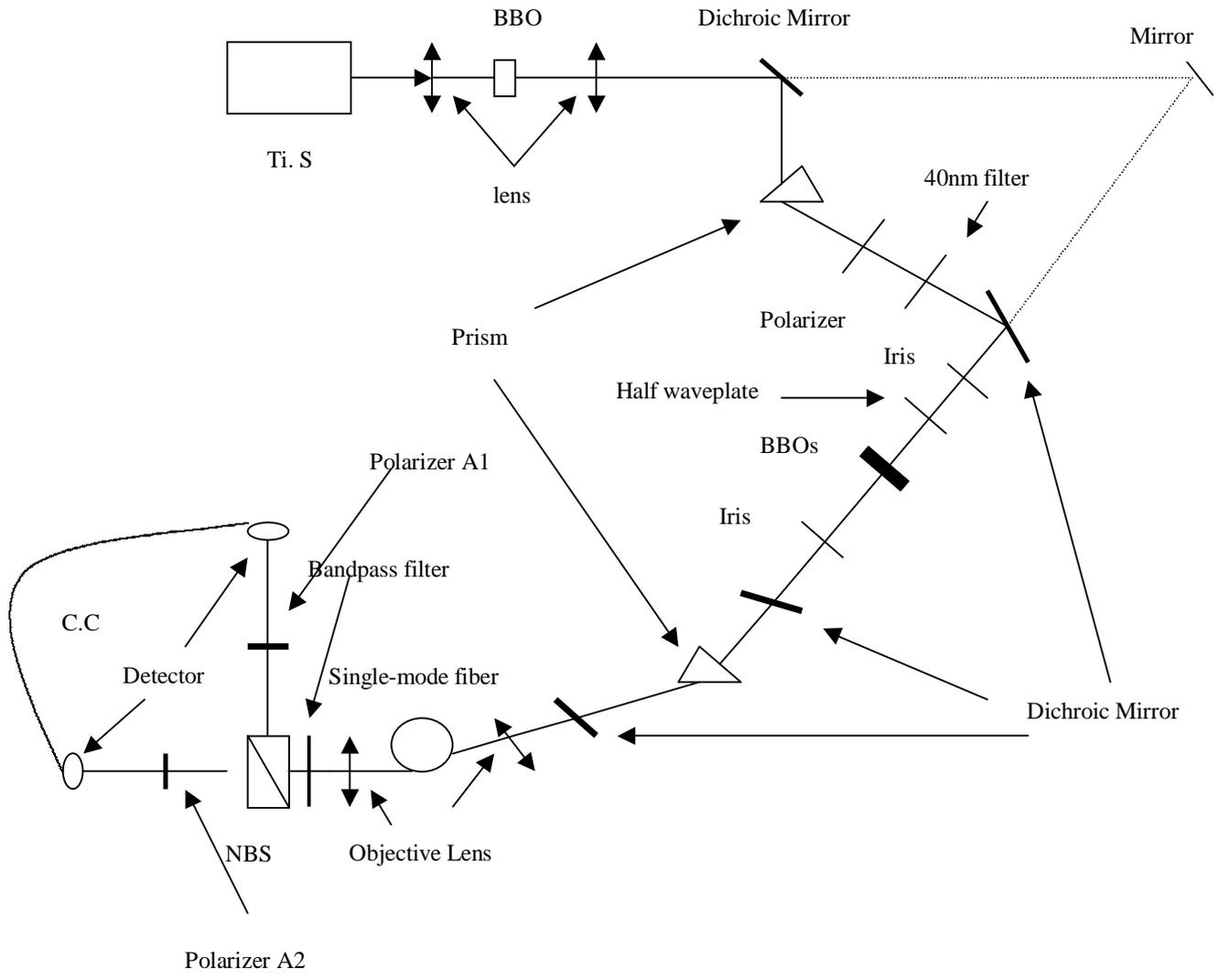
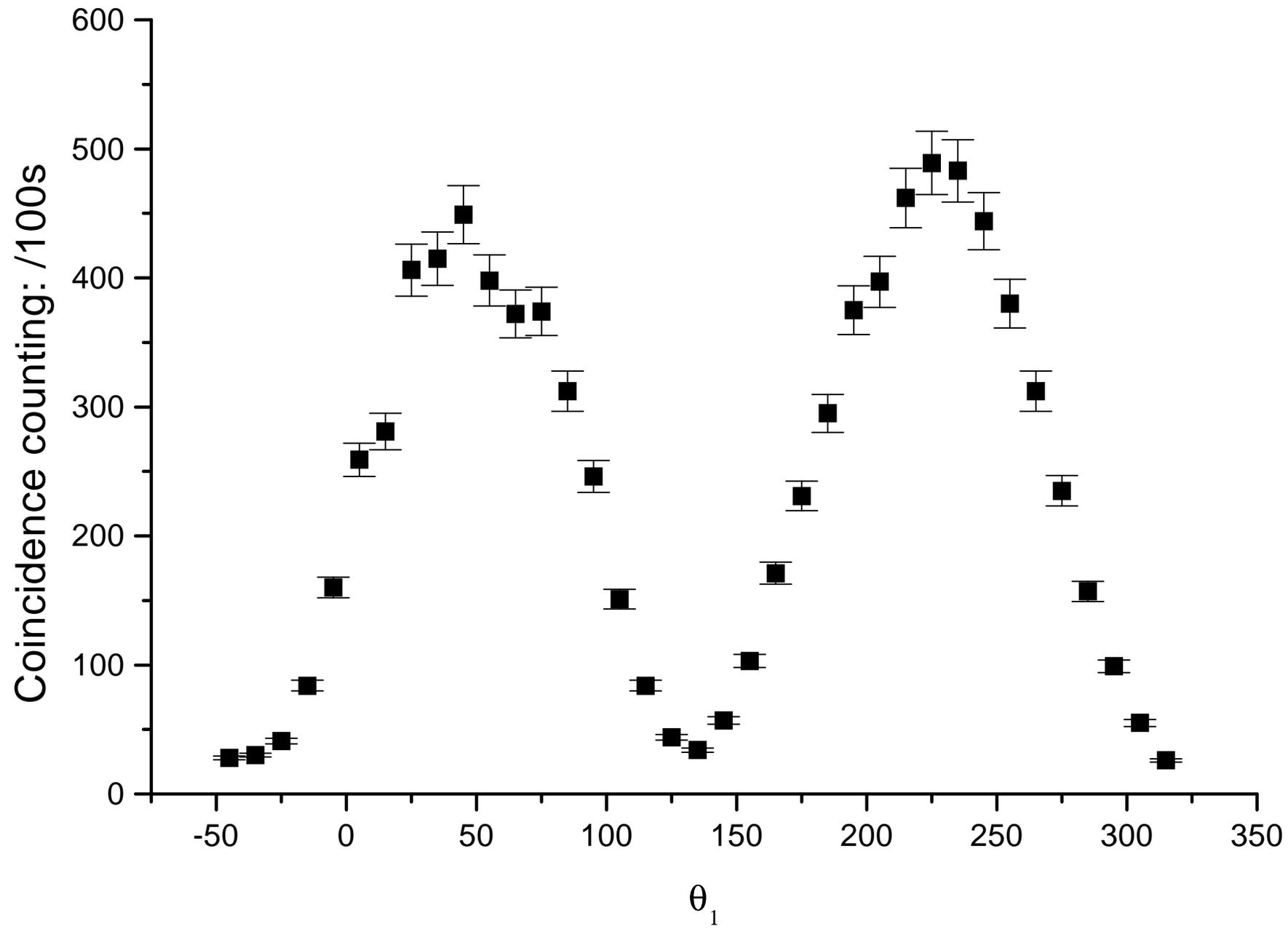
\end{document}